\documentstyle[multicol,aps,psfig]{revtex}

\begin{document}

\title{Jet Tomography of Quark Gluon Plasma}
\author{Xin-Nian Wang} 
\address{
Nuclear Science Division, MS 70R0319,\\
Lawrence Berkeley National Laboratory, Berkeley, CA 94720}

%\date{\today}
\maketitle

\begin{abstract}
Recent experimental measurements of high $p_T$ hadron
spectra and jet correlation at RHIC are analyzed within a parton
model which incoporates initial jet production and final
propagation in heavy-ion collisions. The suppresion of single hadron
spectra, back-to-back correlation, their centrality dependence and
azimuthal anisotropy point to a dense matter with an initial parton
density about 30 times of that in a cold heavy nucleus.
\end{abstract}

%\pacs{74.25.Fy, 72.15.Jf, 75.30.Kz}

% PACS, the Physics and Astronomy Classification Scheme.

\begin{multicols}{2}

\section{Introduction}
In high-energy heavy-ion collisions, a dense medium of quarks and gluons
is expected to be produced and possibly a quark-gluon plasma is formed.
One important step in establishing evidence of QGP formation is to
charaterize the properties of the dense medium produced, for example, 
the parton and energy density and color confinement, among many other
charateristics. 
Conventionally, one can study the properties of a medium via scattering
experiments with particle beams. In deeply inelastic scattering (DIS)
experiments, for example, leptons scatter off the nucleon medium via
photon exchange with quarks. The response function or the correlation 
function of the electromagnetic currents,
\begin{equation}
W_{\mu\nu}(q)=\frac{1}{4\pi}\int d^4x e^{iq\cdot x}\langle A\mid j^{em}_\mu(0)
j^{em}_\nu(x)\mid A \rangle
\end{equation}
is a direct measurement of the quark distributions in a nucleon or nucleus.
Such experiments have provided unprecedented information about partonic
structure of nucleons and nucleus and confirmed the prediction of 
QCD evolution\cite{Martin:2001es}.

For dynamic systems such as that produced in heavy-ion collisions, one
can no longer use the technique of scattering with a beam of particles
because of the transient nature of the matter.
The lifetime of the system is very short, on the order of a 
few fm/$c$. The initial spatial size is only the size of the heaviest 
nuclei, about 6 fm in diameter in the transverse
dimension. The system expands very rapidly both in the 
longitudinal and transverse direction. These characteristics 
make it impossible to use external probes to study the properties 
of the produced dense matter in high-energy heavy-ion
collisions. Fortunately, one can
prove that the thermal average of the above
correlation function gives the photon emission rate from the evolving
system. The emission rate depends mainly on the local temperature or the
parton density while the total yield also depends the whole evolution 
history of the system. Therefore, a strongly interacting system can reveal
its properties and dynamics through photon and dilepton emission. One can
further study the resonance properties of the emitted virtual photons and
their medium modification. The screening of strong interaction in
a color deconfined medium should lead to dissociation of the binding
states and thus the quarkonia suppression \cite{Matsui:1986dk}. Such
color screening is a result of strong interaction between quarks and
gluons at high density and temperature. The same interaction will also
cause attenuation of fast and energetic partons propagating through the
medium. Such an effect is the underlying physics of the 
jet quenching \cite{wg92}
phenomenon and jet tomography technique for studying properties of dense
matter in high-energy heavy-ion collisions.

Jet quenching as a probe of the dense matter in heavy-ion collisions,
takes advantage of the hard processes of jet production in high-energy 
heavy-ion collisions. Similar to the technology of computed
tomography (CT), study of these energetic particles,
their initial production and interaction with the dense medium, can yield
critical information about the properties of the matter that is otherwise
difficult to access through soft hadrons from the hadronization of the
bulk medium. Though relatively rare with small cross
sections, the jet production rate can be calculated perturbative in QCD
and agrees well with experimental measurements in high-energy
$pp(\bar{p})$ collisions. A critical component of the jet tomography is
then to understand the jet attenuation through dense matter as it
propagates through the medium.

\section{Modified Fragmentation Function}
 
A direct manifest of jet quenching is the modification of the 
fragmentation function of the produced parton, 
$D_{a\rightarrow h}(z,\mu^2)$ which can be measured
directly. This modification can be directly translated into the 
energy loss of the leading parton.

To demonstrate medium modified fragmentation function and
parton energy loss, one can study deeply inelastic scattering (DIS)
 $eA$ \cite{Wang:2001if,Guo:2000nz,bwzxnw}. Here,
we consider the semi-inclusive processes,
$e(L_1) + A(p) \longrightarrow e(L_2) + h (\ell_h) +X$,
where $L_1$ and $L_2$ are the four-momenta of the incoming and the
outgoing leptons, and $\ell_h$ is the observed hadron momentum.
The differential
cross section for the semi-inclusive process can be expressed as
\begin{equation}
E_{L_2}E_{\ell_h}\frac{d\sigma_{\rm DIS}^h}{d^3L_2d^3\ell_h}
=\frac{\alpha^2_{\rm EM}}{2\pi s}\frac{1}{Q^4} L_{\mu\nu}
E_{\ell_h}\frac{dW^{\mu\nu}}{d^3\ell_h} \; ,
\label{sigma}
\end{equation}
where $p = [p^+,0,{\bf 0}_\perp] \label{eq:frame}$
is the momentum per nucleon in the nucleus,
$q =L_2-L_1 = [-Q^2/2q^-, q^-, {\bf 0}_\perp]$ the momentum transfer,
$s=(p+L_1)^2$ and $\alpha_{\rm EM}$ is the electromagnetic (EM)
coupling constant. $L_{\mu\nu}$ is the leptonic tensor
while $W_{\mu\nu}$ is the semi-inclusive hadronic tensor.

In the parton model with the collinear factorization approximation,
the leading-twist contribution to the semi-inclusive cross section
can be factorized into a product of parton distributions,
parton fragmentation functions and the hard partonic cross section.
Including all leading log radiative corrections, the lowest order
contribution from a single
hard $\gamma^*+ q$ scattering can be written as
\begin{eqnarray}
\frac{dW^S_{\mu\nu}}{dz_h}
&= &\sum_q e_q^2 \int dx f_q^A(x,\mu_I^2) H^{(0)}_{\mu\nu}(x,p,q)
\nonumber \\
&\times& D_{q\rightarrow h}(z_h,\mu^2)\, ,\label{Dq}
\end{eqnarray}
where $H^{(0)}_{\mu\nu}(x,p,q)$ is the hard part of the
process in leading order, the momentum fraction carried by 
the hadron is defined as
$z_h=\ell_h^-/q^-$ and $x_B=Q^2/2p^+q^-$ is the Bjorken variable.
$\mu_I^2$ and $\mu^2$ are the factorization scales for the initial
quark distributions $f_q^A(x,\mu_I^2)$ in a nucleus and the fragmentation
functions $D_{q\rightarrow h}(z_h,\mu^2)$, respectively.

In a nuclear medium, the propagating quark in DIS will experience additional
scatterings with other partons from the nucleus. The rescatterings may
induce additional gluon radiation and cause the leading quark to lose
energy. Such induced gluon radiations will effectively give rise to
additional terms in the evolution equation leading to the modification of the
fragmentation functions in a medium. These are the so-called higher-twist
corrections since they involve higher-twist parton matrix elements and
are power-suppressed. We will consider those contributions that
involve two-parton correlations from two different nucleons inside
the nucleus.

One can apply the generalized factorization  to these multiple scattering
processes\cite{LQS}. In this approximation, the double scattering
contribution to radiative correction can be calculated and the
effective modified fragmentation function is
\begin{eqnarray}
\widetilde{D}_{q\rightarrow h}(z_h,\mu^2)&\equiv&
D_{q\rightarrow h}(z_h,\mu^2)
+\int_0^{\mu^2} \frac{d\ell_T^2}{\ell_T^2}
\frac{\alpha_s}{2\pi} \int_{z_h}^1 \frac{dz}{z} \nonumber \\
& & \left[ \Delta\gamma_{q\rightarrow qg}(z,x,x_L,\ell_T^2)\right.
 D_{q\rightarrow h}(z_h/z)  \nonumber \\
&+& \left. \Delta\gamma_{q\rightarrow gq}(z,x,x_L,\ell_T^2)
D_{g\rightarrow h}(z_h/z)\right] \, , \label{eq:MDq}
\end{eqnarray}
where $D_{q\rightarrow h}(z_h,\mu^2)$ and
$D_{g\rightarrow h}(z_h,\mu^2)$ are the leading-twist
fragmentation functions. The modified splitting functions are
given as
\begin{eqnarray}
\Delta\gamma_{q\rightarrow qg}(z,x,x_L,\ell_T^2)&=&
\left[\frac{1+z^2}{(1-z)_+}T^{A}_{qg}(x,x_L) \right. \nonumber \\
& & \hspace{-0.8in}+\left. \delta(1-z)\Delta T^{A}_{qg}(x,\ell_T^2) \right]
\frac{2\pi\alpha_s C_A}
{\ell_T^2 N_c f_q^A(x,\mu_I^2)}\, ,
\label{eq:r1}\\
\Delta\gamma_{q\rightarrow gq}(z,x,x_L,\ell_T^2)
&=& \Delta\gamma_{q\rightarrow qg}(1-z,x,x_L,\ell_T^2). \label{eq:r2}
\end{eqnarray}
Here, the fractional momentum is defined as
$x_L =\ell_T^2/2p^+q^-z(1-z)$ and $x=x_B=Q^2/2p^+q^-$ is 
the Bjorken variable. The twist-four parton matrix elements of the nucleus, 
\begin{eqnarray}
T^{A}_{qg}(x,x_L)&=& \int \frac{dy^{-}}{2\pi}\, dy_1^-dy_2^-
e^{i(x+x_L)p^+y^-}(1-e^{-ix_Lp^+y_2^-})\nonumber  \\
& & \hspace{-0.4in}\times (1-e^{-ix_Lp^+(y^--y_1^-)})
\theta(-y_2^-)\theta(y^- -y_1^-)\nonumber  \\
&&\hspace{-0.4in}\times\frac{1}{2}\langle A | \bar{\psi}_q(0)\,
\gamma^+\, F_{\sigma}^{\ +}(y_{2}^{-})\, F^{+\sigma}(y_1^{-})\,\psi_q(y^{-})
| A\rangle  \;\; , \label{Tqg}
\end{eqnarray}
has a dipole-like structure which is a result of Landau-Pomerachuck-Migdal
(LPM) interference in gluon bremsstrahlung. Here the intrinsic transverse
momentum is neglected. In the limit of collinear
radiation ($x_L\rightarrow 0$) or when the formation time of the
gluon radiation, $\tau_f\equiv 1/x_Lp^+$, is much larger
than the nuclear size, the destructive interference leads 
to the LPM interference effect.

Using the factorization 
approximation\cite{Wang:2001if,Guo:2000nz,LQS,ow}, we can
relate the twist-four parton matrix elements of the nucleus
to the twist-two parton distributions of nucleons and the
nucleus,
\begin{equation}
T^{A}_{qg}(x,x_L)\approx \frac{\widetilde{C}}{x_A}
(1-e^{-x_L^2/x_A^2}) f_q^A(x),
\label{modT2}
\end{equation}
where $\widetilde{C}\equiv 2C x_Tf^N_g(x_T)$ is considered a constant.
One can identify $1/x_Lp^+=2q^-z(1-z)/\ell_T^2$ 
as the formation time of the emitted gluons. When it becomes comparable
or larger than the nuclear size, the above matrix element vanishes,
demonstrating a typical LPM interference effect.

Since the LPM interference suppresses 
gluon radiation whose formation time ($\tau_f \sim Q^2/\ell_T^2p^+$)
is larger than the nuclear 
size $MR_A/p^+$ in our chosen frame, $\ell_T^2$ should then have a 
minimum value of $\ell_T^2\sim Q^2/MR_A\sim Q^2/A^{1/3}$. 
Here $M$ is the nucleon mass.
Therefore, the leading higher-twist
contribution is 
proportional to $\alpha_s R_A/\ell_T^2 \sim \alpha_s R_A^2/Q^2$
due to double scattering and depends quadratically on the nuclear size $R_A$.

\begin{figure}[]
\centerline{\psfig{figure=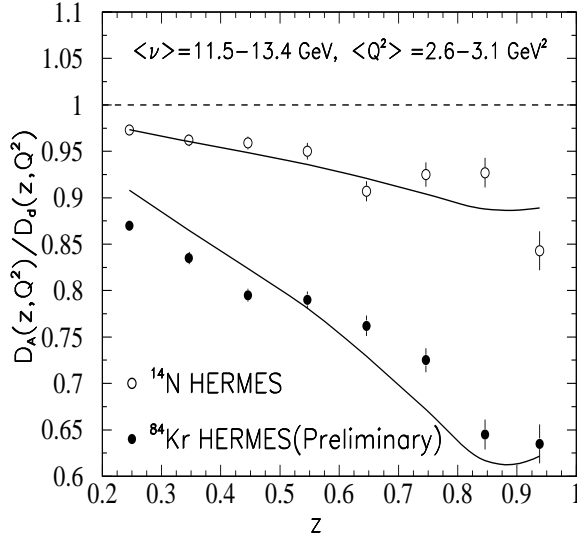,width=3.0in,height=2.8in}}
\baselineskip=10pt
\caption{Predicted nuclear modification of jet fragmentation function
is compared to the HERMES data \protect\cite{hermes} on ratios of
hadron distributions between $A$ and $d$ targets in DIS.
\label{fig:hermes1}}
\end{figure}

With the assumption of the factorized form 
of the twist-4 nuclear parton matrices, there is only one free 
parameter $\widetilde{C}(Q^2)$
which represents quark-gluon correlation strength inside nuclei.
Once it is fixed, one can predict the $z$, energy and
nuclear dependence of the medium modification of the fragmentation
function. Shown in Fig.~\ref{fig:hermes1} are the calculated 
nuclear modification factor of the fragmentation functions for $^{14}N$ 
and $^{84}Kr$ targets as compared to the recent HERMES data\cite{hermes}.
The predicted shape of the $z$-dependence agrees well 
with the experimental data.  A remarkable feature of the prediction
is the quadratic $A^{2/3}$ nuclear size dependence, which is verified 
for the first time by an experiment.
By fitting the overall suppression for one nuclear target, 
we obtain the only parameter in our calculation,
$\widetilde{C}(Q^2)=0.0060$ GeV$^2$ 
with $\alpha_{\rm s}(Q^2)=0.33$ at $Q^2\approx 3$ GeV$^2$.
The predicted $\nu$-dependence also agrees with the experimental
data \cite{ww02}. 

We can quantify the modification of the fragmentation
by the quark energy loss which is defined as the momentum fraction
carried by the radiated gluon,
\begin{eqnarray}
\langle\Delta z_g\rangle
&=& \int_0^{\mu^2}\frac{d\ell_T^2}{\ell_T^2}
\int_0^1 dz \frac{\alpha_s}{2\pi}
 z\,\Delta\gamma_{q\rightarrow gq}(z,x_B,x_L,\ell_T^2) \nonumber \\
&=&\widetilde{C}\frac{C_A\alpha_s^2}{N_c}
\frac{x_B}{x_AQ^2} \int_0^1 dz \frac{1+(1-z)^2}{z(1-z)}\nonumber \\
&\times&\int_0^{x_\mu} \frac{dx_L}{x_L^2}(1-e^{-x_L^2/x_A^2}),
\label{eq:heli-loss}
\end{eqnarray}
where $x_\mu=\mu^2/2p^+q^-z(1-z)=x_B/z(1-z)$ if we choose the
factorization scale as $\mu^2=Q^2$.
When $x_A\ll x_B\ll 1$ we can estimate the leading quark energy
loss roughly as
\begin{eqnarray}
\langle \Delta z_g\rangle(x_B,\mu^2)& \approx &
\widetilde{C}\frac{C_A\alpha_s^2}{N_c}\frac{x_B}{Q^2
x_A^2}6\sqrt{\pi}\ln\frac{1}{2x_B}\, .
\label{eq:appr1-loss}
\end{eqnarray}
Since $x_A=1/MR_A$, the energy loss $\langle \Delta
z_g\rangle$ thus depends quadratically on the nuclear size.

In the rest frame of the nucleus, $p^+=m_N$, $q^-=\nu$, and
$x_B\equiv Q^2/2p^+q^-=Q^2/2m_N\nu$. One can
get the averaged total energy loss as
$ \Delta E=\nu\langle\Delta z_g\rangle
\approx  \widetilde{C}(Q^2)\alpha_{\rm s}^2(Q^2)
m_NR_A^2(C_A/N_c) 3\ln(1/2x_B)$.
With the determined value of $\widetilde{C}$, 
$\langle x_B\rangle \approx 0.124$ in the HERMES experiment\cite{hermes}
and the average distance $\langle L_A\rangle=R_A\sqrt{2/\pi}$
for the assumed Gaussian nuclear distribution,
one gets the quark energy 
loss $dE/dL\approx 0.5$ GeV/fm inside a $Au$ nucleus.

\section{Jet Quenching in Hot Medium at RHIC}

To extend our study of modified fragmentation functions to 
jets in heavy-ion collisions, we can
assume $\langle k_T^2\rangle\approx \mu^2$ (the Debye screening mass)
and a gluon density profile
$\rho(y)=(\tau_0/\tau)\theta(R_A-y)\rho_0$ for a 1-dimensional 
expanding system. Since the initial jet production 
rate is independent of the final gluon density which can be 
related to the parton-gluon scattering cross 
section\cite{Baier:1996sk} 
[$\alpha_s x_TG(x_T)\sim \mu^2\sigma_g$], one has then
\begin{equation}
\frac{\alpha_s T_{qg}^A(x_B,x_L)}{f_q^A(x_B)} \sim
\mu^2\int dy \sigma _g \rho(y)
[1-\cos(y/\tau_f)],
\end{equation}
where $\tau_f=2Ez(1-z)/\ell_T^2$ is the gluon formation time. One
can recover the form of energy loss in a thin plasma obtained 
in the opacity expansion approach\cite{Gyulassy:2000gk},
\begin{eqnarray}
\langle\Delta z_g\rangle &=&\frac{C_A\alpha_s}{\pi}
\int_0^1 dz \int_0^{\frac{Q^2}{\mu^2}}du \frac{1+(1-z)^2}{u(1+u)}
\int_{\tau_0}^{R_A} d\tau\sigma_g\rho(\tau) \nonumber \\
&\times&
\left[1-\cos\left(\frac{(\tau-\tau_0)\,u\,\mu^2}{2Ez(1-z)}\right)\right].
\end{eqnarray}
Keeping only the dominant contribution and assuming 
$\sigma_g\approx C_a 2\pi\alpha_s^2/\mu^2$ ($C_a$=1 for $qg$ and 9/4 for
$gg$ scattering), one obtains the averaged energy loss,
\begin{equation}
\langle \frac{dE}{dL}\rangle \approx \frac{\pi C_aC_A\alpha_s^3}{R_A}
\int_{\tau_0}^{R_A} d\tau \rho(\tau) (\tau-\tau_0)\ln\frac{2E}{\tau\mu^2}.
\label{effloss}
\end{equation}
Neglecting the logarithmic dependence on $\tau$, the averaged energy loss
in a 1-dimensional expanding system can be expressed as
$\langle\frac{dE}{dL}\rangle_{1d} \approx (dE_0/dL) (2\tau_0/R_A)$,
where $dE_0/dL\propto \rho_0R_A$
is the energy loss in a static medium with the same gluon density $\rho_0$ 
as in a 1-d expanding system at time $\tau_0$.
Because of the expansion, the averaged energy loss $\langle dE/dL\rangle_{1d}$
is suppressed as compared to the static case and does not depend linearly
on the system size.

In order to calculate the effects of parton energy loss on the
attenuation pattern of high $p_T$ partons in nuclear collisions, 
we use a simpler effective modified fragmentation
function\cite{Wang:1996yh,Wang:1996pe},
\begin{eqnarray}
D_{h/c}^\prime(z_c,Q^2,\Delta E_c) 
&=&e^{-\langle\frac{\Delta L}{\lambda}\rangle}D^0_{h/c}(z_c,Q^2)
+(1-e^{-\langle \frac{\Delta L}{\lambda}\rangle})\nonumber \\
&& \hspace{-0.8in}
\times\left[ \frac{z_c^\prime}{z_c} D^0_{h/c}(z_c^\prime,Q^2) 
+\langle \frac{\Delta L}{\lambda}\rangle
\frac{z_g^\prime}{z_c} D^0_{h/g}(z_g^\prime,Q^2)\right]
\label{modfrag} 
\end{eqnarray}
where $z_c^\prime,z_g^\prime$ are the rescaled momentum fractions.
This effective model is found to reproduce the pQCD result 
from Eq.(\ref{eq:MDq}) very well, but only when
$\Delta z=\Delta E_c/E$ is set to
 be $\Delta z\approx 0.6 \langle z_g\rangle$.
Therefore the actual averaged parton energy loss should be
$\Delta E/E=1.6\Delta z$ with $\Delta z$ extracted from the 
effective model. The factor 1.6 is mainly
caused by the unitarity correction effect in 
the pQCD calculation.

Since gluons are bosons, there should also
be stimulated gluon emission and absorption by the propagating parton
because of the presence of thermal gluons in the hot medium.  Such detailed
balance is crucial for parton thermalization and should also be important
for calculating the energy loss of an energetic parton in a hot 
medium\cite{ww01}. Taking into account such detailed balance in
gluon emission, one can then get the
asymptotic behavior of the effective energy loss
in the opacity expansion framework \cite{ww01},
\begin{eqnarray}
   {\Delta E\over E}\approx &&
   {{\alpha_s C_F \mu^2 L^2}\over 4\lambda_gE}
   \left[\ln{2E\over \mu^2L} -0.048\right]\nonumber \\
   &&\hspace{-0.3in}-
   {{\pi\alpha_s C_F}\over 3} {{LT^2}\over {\lambda_g E^2}}
   \left[
   \ln{{\mu^2L}\over T} -1+\gamma_{\rm E}-{{6\zeta^\prime(2)}\over\pi^2}
\right],
 \end{eqnarray}
where the first term is from the induced bremsstralung and the second
term is due to gluon absorption in detailed balance which effectively
reduce the total parton energy loss in the medium.
Numerical calculations show that the effect of the gluon
absorption is small and can be neglected for partons with 
very high energy. 
However, the thermal absorption reduces the effective
parton energy loss by about 30-10\% for intermediate values of
parton energy. This will increase the energy dependence of the
effective parton energy loss in the intermediate energy region.
One can parameterize such energy dependence as,
\begin{equation}
 \langle\frac{dE}{dL}\rangle_{1d}=\epsilon_0 (E/\mu-1.6)^{1.2}
 /(7.5+E/\mu),
\label{eq:loss}
\end{equation}
The threshold is the consequence of gluon absorption that competes
with radiation that effectively shuts off the energy loss. The
parameter $\mu$ is set to be 1 GeV in the calculation.

To calculate the modified high $p_T$ spectra in $A+A$ collisions,
we use a LO pQCD model \cite{Wang:1998ww,Wang:2003mm},
\begin{eqnarray}
  \frac{d\sigma^h_{AA}}{dyd^2p_T}&=&K\sum_{abcd} 
  \int d^2b d^2r dx_a dx_b d^2k_{aT} d^2k_{bT}  \nonumber \\
  &\times& t_A(r)t_A(|{\bf b}-{\bf r}|) 
  g_A(k_{aT},r)  g_A(k_{bT},|{\bf b}-{\bf r}|) \nonumber \\
  &\times& f_{a/A}(x_a,Q^2,r)f_{b/A}(x_b,Q^2,|{\bf b}-{\bf r}|) \nonumber \\
 &\times& \frac{D_{h/c}^\prime (z_c,Q^2,\Delta E_c)}{\pi z_c}  
  \frac{d\sigma}{d\hat{t}}(ab\rightarrow cd), \label{eq:nch_AA}
\end{eqnarray}
with medium modified fragmentation funcitons $D_{h/c}^\prime$
given by Eq.~(\ref{modfrag}) and the fragmentation functions in 
free space $D^0_{h/c}(z_c,Q^2)$ are given by the BBK 
parameterization \cite{bkk}.
Here, $z_c=p_T/p_{Tc}$, $y=y_c$, $\sigma(ab\rightarrow cd)$ are 
elementary parton scattering cross sections and $t_A(b)$ is the 
nuclear thickness function normalized to $\int d^2b t_A(b)=A$. 
We will use a hard-sphere model of nuclear distribution in this paper.
The $K\approx 1.5-2$ factor is used to account for higher order pQCD 
corrections.

\begin{figure}
\centerline{\psfig{figure=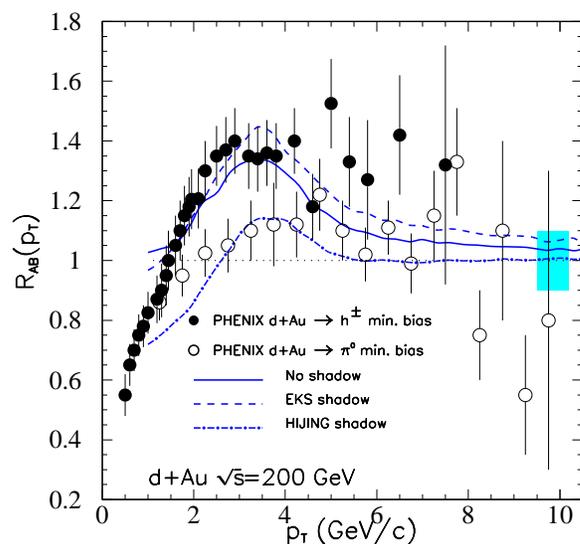,width=3.0in,height=2.8in}}
\baselineskip=10pt
\caption{The first predictions \protect\cite{Wang:1998ww} of
the Cronin effect in $p+Au$ collisions 
at $\sqrt{s}=200$ GeV are compared to the recent RHIC data from 
PHENIX \protect\cite{dauphenix} and STAR \protect\cite{daustar}.
\label{fig:dau}}
\end{figure}

The parton distributions per nucleon $f_{a/A}(x_a,Q^2,r)$
inside the nucleus are assumed to be factorizable into the parton 
distributions in a free nucleon given by the MRSD$-^{\prime}$  
parameterization \cite{Martin} and the impact-parameter dependent 
nuclear modification factor which will given by the new 
HIJING parameterization \cite{lw02}. The initial transverse momentum
distribution $g_A(k_T,Q^2,b)$ is assumed to have a Gaussian form
with a width that includes both an intrinsic part in a nucleon and 
nuclear broadening. This model has been fitted to the nuclear
modification of the $p_T$ spectra in $p+A$ collisions at up
to the Fermilab energy $\sqrt{s}=40$ GeV. Shown in Fig.~\ref{fig:dau}
are the first prediction made in 1998 \cite{Wang:1998ww}
of the Cronin effect at RHIC 
for $p+Au$ collisions at $\sqrt{s}=200$ GeV as compared to
the recent RHIC data. As one can see, the initial multiple
scattering in nuclei can give some moderate Cronin enhancement
of the high $p_T$ spectra. Therefore, any suppression of the
high $p_T$ spectra in $Au+Au$ collisions has to be caused by
jet quenching.

We assume a 1-dimensional expanding medium with a gluon 
density $\rho_g(\tau,r)$ that is proportional to the 
transverse profile of participant nucleons.
According to Eq.~\ref{effloss}, we will calculate impact-parameter
dependence of the energy loss as
\begin{equation}
\Delta E(b,r,\phi)\approx \langle\frac{dE}{dL}\rangle_{1d}
\int_{\tau_0}^{\Delta L} d\tau\frac{\tau-\tau_0}{\tau_0\rho_0}
\rho_g(\tau,b,\vec{r}+\vec{n}\tau),
\end{equation}
where $\Delta L(b,\vec{r},\phi)$ is the distance a jet, produced at
$\vec{r}$, has to travel along $\vec{n}$ at an azimuthal 
angle $\phi$ relative to the reaction plane in a collision 
with impact-parameter $b$. Here, $\rho_0$ is the averaged 
initial gluon density at $\tau_0$ 
in a central collision and $\langle dE/dL\rangle_{1d}$ 
is the average parton energy loss over a distance $R_A$
in a 1-d expanding medium with an initial uniform gluon 
density $\rho_0$. The corresponding energy loss 
in a static medium with a uniform gluon density 
$\rho_0$ over a distance $R_A$ is \cite{ww02}
$dE_0/dL=(R_A/2\tau_0)\langle dE/dL\rangle_{1d}$.
We will use the parameterization in Eq.~(\ref{eq:loss})
for the effective energy dependence of the parton
quark energy loss.

\begin{figure}
\centerline{\psfig{figure=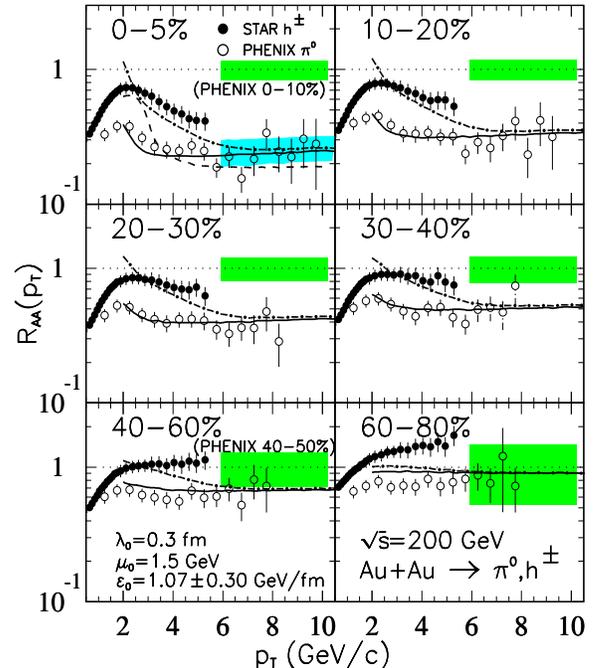,width=3.0in,height=3.5in}}
%\centerline{\includegraphics[angle=-90,width=8.4cm]{r_aacern.eps}
\caption{Hadron suppression factors in $Au+Au$ collisions
as compared to data from STAR\protect\cite{star-r} and 
PHENIX \protect\cite{phenix-r}. See text for a detailed explanation.
\label{fig1}}
\end{figure}

Shown in Fig.~\ref{fig1} are the calculated nuclear 
modification factors
$R_{AB}(p_T)=d\sigma^h_{AB}/\langle N_{\rm binary}\rangle d\sigma^h_{pp}$
for hadron spectra ($|y|<0.5$) in $Au+Au$ collisions 
at $\sqrt{s}=200$ GeV, as compared to experimental 
data \cite{star-r,phenix-r}. Here,
$\langle N_{\rm binary}\rangle=\int d^2bd^2r t_A(r)t_A(|\vec{b}-\vec{r}|)$.
To fit the observed $\pi^0$ suppression (solid lines) in the most 
central collisions, we have used $\mu=1.5$ GeV,
$\epsilon_0=1.07$ GeV/fm and $\lambda_0=1/(\sigma\rho_0)=0.3$ fm.
The hatched area (also in other figures in this paper) indicates 
a variation of $\epsilon_0=\pm 0.3$ GeV/fm.
The hatched boxes around $R_{AB}=1$ represent experimental
errors in overall normalization.
Nuclear $k_T$ broadening and parton shadowing together give a slight 
enhancement of hadron spectra at intermediate $p_T=2-4$ GeV/$c$ 
without parton energy loss.

The flat $p_T$ dependence of the $\pi^0$ suppression is 
a consequence of the strong energy dependence of the
parton energy loss. The slight rise of $R_{AB}$
at $p_T<4$ GeV/$c$ in the calculation is due to the detailed
balance effect in the effective parton energy loss. In this
region, one expects the fragmentation picture to gradually 
lose its validity and is taken over by other non-perturbative 
effects, especially for kaons and baryons.
As a consequence, the $(K+p)/\pi$ ratio in central $Au+Au$
collisions is significantly larger than in peripheral $Au+Au$ or
$p+p$ collisions. To take into account this effect, 
we add a nuclear dependent (proportional to
$\langle N_{\rm binary}\rangle$) soft
component to kaon and baryon fragmentation functions so that
$(K+p)/\pi\approx 2$ at $p_T\sim 3$ GeV/$c$ in the most 
central $Au+Au$ collisions and approaches its $p+p$ value 
at $p_T>5$ GeV/$c$. The resultant suppression for
total charged hadrons (dot-dashed) and the centrality dependence 
agree well with the STAR data. One can directly relate $h^{\pm}$ 
and $\pi^0$ suppression via the $(K+p)/\pi$ ratio: 
$R_{AA}^{h^{\pm}}=R_{AA}^{\pi^0}[1+(K+p)/\pi]_{AA}/[1+(K+p)/\pi]_{pp}$.
It is clear from the data that $(K+p)/\pi$ becomes the same for
$Au+Au$ and $p+p$ collisions at $p_T>5$ GeV/$c$.
To demonstrate the sensitivity to the parameterized
parton energy loss in the intermediate $p_T$ region, 
we also show $R_{AA}^{h^{\pm}}$ in 0-5\% centrality (dashed line)
for $\mu=2.0$ GeV and $\epsilon_0=2.04$ GeV/fm without the 
soft component.

In the same LO pQCD parton model, one can also calculate di-hadron
spectra,
\begin{eqnarray}
  &&E_1E_2\frac{d\sigma^{h_1h_2}_{AA}}{d^3p_1d^3p_2}=\frac{K}{2}\sum_{abcd} 
  \int d^2b d^2r dx_a dx_b d^2k_{aT} d^2k_{bT}  \nonumber \\
 &\times& t_A(r)t_A(|{\bf b}-{\bf r}|) 
  g_A(k_{aT},r)  g_A(k_{bT},|{\bf b}-{\bf r}|) \nonumber \\
   &\times& f_{a/A}(x_a,Q^2,r)  f_{b/A}(x_b,Q^2,|{\bf b}-{\bf r}|)  
   \nonumber \\
  &\times& D_{h/c}(z_c,Q^2,\Delta E_c)
  D_{h/d}(z_d,Q^2,\Delta E_d) 
 \frac{\hat{s}}{2\pi z_c^2 z_d^2} \nonumber \\
&\times& \frac{d\sigma}{d\hat{t}}(ab\rightarrow cd)
 \delta^4(p_a+p_b-p_c-p_d),
 \label{eq:dih}
\end{eqnarray}
for two back-to-back hadrons from independent fragmentation
of the back-to-back jets. 
Let us assume hadron $h_1$ is a triggered hadron 
with $p_{T1}=p_T^{\rm trig}$. One can define a hadron-triggered 
fragmentation function (FF) 
as the back-to-back correlation with respect to the triggered hadron:
\begin{equation}
  D^{h_1h_2}(z_T,\phi,p^{\rm trig}_T)=
  p^{\rm trig}_T \frac{d\sigma^{h_1h_2}_{AA}/d^2p^{\rm trig}_T dp_Td\phi}
  {d\sigma^{h_1}_{AA}/d^2p^{\rm trig}_T},
\end{equation}
similarly to the direct-photon triggered FF \cite{Wang:1996yh,Wang:1996pe} 
in $\gamma$-jet events. Here, $z_T=p_T/p^{\rm trig}_T$ and 
integration over $|y_{1,2}|<\Delta y$ is implied. 
In a simple parton model, the two jets should be
exactly back-to-back. The initial parton transverse momentum distribution
in our model will give rise to a Gaussian-like angular distribution.
In addition, we also take into account transverse momentum smearing
within a jet using a Gaussian distribution with a width of
$\langle k_\perp\rangle=0.6$ GeV/$c$. Hadrons from 
the soft component are assumed to be uncorrelated.

\begin{figure}
\centerline{\psfig{figure=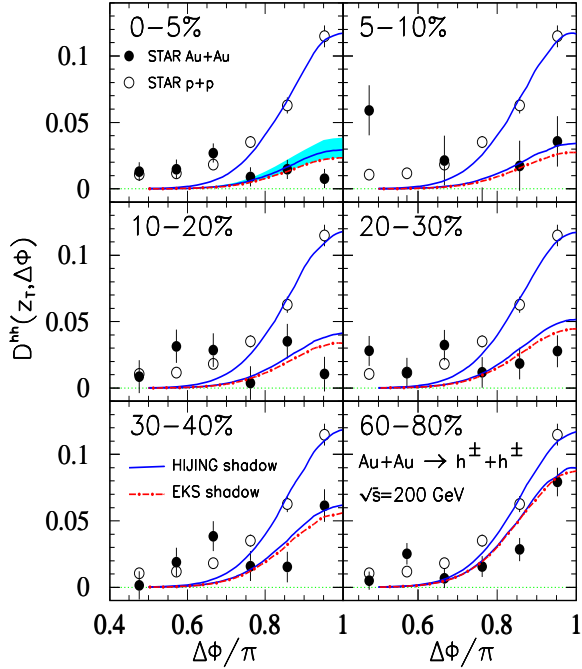,width=3.0in,height=3.5in}}
%\centerline{\includegraphics[angle=-90,width=8.4cm]{phicern.eps}
\caption{Back-to-back correlations for charged hadrons 
with $p^{\rm trig}_T>p_T>2$ GeV/$c$, 
$p^{\rm trig}_T=4-6$ GeV/$c$ and $|y|<0.7$ in $Au+Au$ (lower curves) 
and $p+p$ (upper curves)
collisions as compared to the STAR\protect\cite{star-c} data.
\label{fig3}}
\end{figure}

Shown in Fig.~\ref{fig3} are the calculated back-to-back correlations 
for charged hadrons in $Au+Au$ collisions as compared to the STAR 
data \cite{star-c}. The same energy loss that is used to calculate 
single hadron suppression and azimuthal anisotropy can also describe
well the observed away-side hadron suppression and its centrality
dependence. In the data, a background 
$B(p_T)[1+2v_2^2(p_T)\cos(2\Delta\phi)]$ from uncorrelated hadrons
and azimuthal anisotropy has been subtracted. The value of $v_2(p_T)$
is measured independently while
$B(p_T)$ is determined by fitting the observed correlation in the
region $0.75<|\phi|<2.24$ rad \cite{star-c}.

With both the single spectra and dihadron spectra, 
the extracted average energy loss in this model calculation
for a 10 GeV quark in the expanding medium is 
$\langle dE/dL\rangle_{1d}\approx 0.85 \pm  0.24$ GeV/fm, which
is equivalent to $dE_0/dL\approx 13.8 \pm 3.9$ GeV/fm in a static and
uniform medium over a distance $R_A=6.5$ fm. This value 
is about a factor of 2 larger than a previous estimate \cite{ww02}
because of the variation of gluon density along the propagation
path and the more precise RHIC data considered .

\section{Summary}

In summary, with the recent measurements of high $p_T$ hadron spectra and
jet correlations in $d+Au$ collisions at RHIC, the observed jet quenching
in $Au+Au$ collisions has been established as a consequence of final state
interaction between jets and the produced dense medium. The collective
body of data: suppression of high $p_T$ spectra and back-to-back jet
correlation, high $p_T$ anisotropy and centrality dependences of the
observables, indicate that the cause of the jet quenching is due to parton
energy loss rather than final state absorption of hadrons.

A simultaneous phenomenological study of the suppression of
hadron spectra and back-to-back correlations, and high-$p_T$ azimuthal
anisotropy in high-energy heavy-ion collisions within a single LO pQCD
parton model incorporating current theoretical understanding
of parton energy loss can describe the experimental data of $Au+Au$ 
collisions very well. With HIJING (EKS) parton shadowing,
the extracted average energy loss for 
a 10 GeV quark in the expanding medium is 
$\langle dE/dL\rangle_{1d}\approx 0.85 (0.99) \pm  0.24$ GeV/fm, which
is equivalent to $dE_0/dL\approx 13.8 (16.1) \pm 3.9$ GeV/fm in a static and
uniform medium over a distance $R_A=6.5$ fm. This value 
is about a factor of 2 larger than a previous estimate \cite{ww02}
because of the variation of gluon density along the propagation
path and the more precise RHIC data considered here .

\section*{Acknowledgement}

This work is supported  by
the Director, Office of Energy
Research, Office of High Energy and Nuclear Physics, Divisions of 
Nuclear Physics, of the U.S. Department of Energy under Contract No.
DE-AC03-76SF00098

\end{multicols}

\end{document}